\documentclass[nolinenumbers]{aa}

\usepackage{graphicx}
\usepackage{txfonts}

\usepackage{hyperref}
\hypersetup{
  colorlinks=true,
  linkcolor=blue,
  citecolor=blue,
  filecolor=magenta,      
  urlcolor=blue,
}

\usepackage{color}
\definecolor{purple2}{RGB}{154,6,128}



\begin{document}

   \title{ALMA Band\,1 observations of the $\rho$\,Oph\,W filament}

   \subtitle{I. Enhanced power from  excess microwave emission  at high spatial frequencies}

   \author{S. Casassus
     \inst{1,2}
     \and
     M. Vidal
     \inst{3}
     \and
     M. C\'arcamo
     \inst{2,4}
     \and
     L. Verstraete
     \inst{5}
     \and
     N. Ysard
     \inst{5,6}
     \and
     E. Habart
     \inst{5}
   }
   
   \institute{Departamento de Astronom\'{\i}a, Universidad de Chile, Casilla 36-D, Santiago, Chile\\
     \email{simon@das.uchile.cl}
     \and
     Data Observatory Foundation, Eliodoro Yáñez 2990, Providencia, Santiago, Chile
\and
     Núcleo de Astroquímica, Facultad de Ingeniería, Universidad Autónoma de Chile, Av. Pedro de Valdivia 425, Providencia, Santiago, Chile
     \and
     University of Santiago of Chile (USACH), Faculty of Engineering, Computer Engineering Department, Chile
     \and
     Institut d’Astrophysique Spatiale (IAS), Universit\'e Paris-Saclay, CNRS, B\^atiment 121, 91405 Orsay Cedex, France
     \and
     Institut de Recherche en Astrophysique et Plan\'etologie (IRAP), Toulouse, France
   }

   \date{Received September 15, 1996; accepted March 16, 1997}

  \abstract
{The $\rho$\,Oph\,W photo-dissociation region (PDR) is an example
    source of bright excess microwave emission (EME), over
     synchrotron, free-free, and the Rayleigh-Jeans tail of the sub-millimetre (sub-mm)
    dust continuum. Its filamentary morphology follows roughly that of
    the IR poly-cyclic aromatic hydrocarbon (PAHs) bands. The EME signal in $\rho$\,Oph\,W 
    drops abruptly above $\sim$30\,GHz and its spectrum can be
    interpreted in terms of electric-dipole radiation from spinning
    dust grains, or ``spinning dust''.  }
{Deep and high-fidelity imaging and spectroscopy of $\rho$\,Oph\,W may reveal the detailed morphology of the EME signal, free from imaging priors, while also enabling  a search for fine structure in its spectrum. The same observations may constrain the spectral index of the high-frequency drop.}
{An ALMA Band\,1 mosaic yields a deep deconvolved image of the filament at 36--44\,GHz, which we  use as template for the  extraction of a spectrum via cross-correlation in the $uv$-plane. Simulations and cross-correlations on near-infrared ancillary data yield estimates of flux-loss and biases.}
{The spectrum is a power law, with no detectable fine structure. It follows a spectral index $\alpha = -0.78\pm0.05$, in frequency,  with some variations along the filament. Interestingly, the Band\,1 power at high spatial frequencies increases relative to that of the IR signal, with  a  factor of two more power in Band\,1 at $\sim 20\arcsec$ than at $\sim 100\,\arcsec$ (relative to IRAC\,3.6\,$\mu$m). An extreme of such radio-only structures is a compact EME source, without  IR counterpart. It is embedded in strong and filamentary Band\,1 signal, while the IRAC maps are smooth in the same region. We provide multi-frequency intensity estimates for spectral modelling.}
{}


\maketitle

\section{Introduction}

Cosmic microwave background (CMB) anisotropy experiments  identified an anomalous diffuse
Galactic foreground in the range 10--90\,GHz \citep[][]{kog96,lei97},
which was confirmed, in particular, by the {\em WMAP}
\citep{Gold2011ApJS..192...15G} and {\em Planck} missions
\citep[][]{Planck2011A&A...536A..20P, Planck2016A&A...594A...1P}. As reviewed in
\citet{Dickinson2018NewAR..80....1D}, this diffuse emission is
correlated with the infrared (IR) thermal emission from dust grains on large
angular scales and at high Galactic latitudes.  The spectral index in
specific intensity ($I_\nu \propto \nu^\alpha$) of the anomalous
Galactic foreground is $\alpha \sim 0$ in the range 15--30\,GHz
\citep[][]{kog96}, but any semblance to optically thin free-free is
dissipated by a drop between 20--40\,GHz, with $\alpha \sim -0.85$ for
high-latitude cirrus \citep[][]{dav06}.

Bright centimetre (cm) wavelength radiation has been reported from a dozen
well-studied molecular clouds, with intensities that are well in excess of the
expected levels for free-free, synchrotron, or the Rayleigh-Jeans tail
of the sub-mm dust emission
\citep[e.g.][]{Finkbeiner2002ApJ...566..898F,
  Watson2005ApJ...624L..89W, Casassus2006ApJ...639..951C,
  Scaife2009MNRAS.394L..46A, Scaife2010, Castellanos2011MNRAS.411.1137C,
  Vidal2011MNRAS.414.2424V, Tibss2012ApJ...754...94T,
  Vidal2020MNRAS.495.1122V, Cepeda-Arroita2021MNRAS.503.2927C}. A common feature of all
cm-bright clouds is that they host conspicuous photo-dissociation
regions (PDRs).  The {\em Planck} mission picked up spectral
variations in this excess microwave emission (EME) from source to
source along the Gould Belt, where the peak frequency is
$\nu_{\rm peak} \sim$26--30\,GHz, while $\nu_{\rm peak} \sim 25\,$GHz
in the diffuse ISM \citep[][]{Planck2013A&A...557A..53P,
  Planck2016A&A...594A..10P}.

The prevailing interpretation for EME, also called anomalous microwave
emission (AME), is electric-dipole radiation from spinning very small
grains, or `spinning dust' \citep[][]{dra98}. A comprehensive review
of all-sky surveys and targeted observations supports this spinning
dust interpretation \citep[][]{Dickinson2018NewAR..80....1D}. The
carriers of spinning dust remain to be identified, however, and could
be poly-cyclic aromatic hydrocarbons (PAHs)
\citep[][]{Ali-Haimoud2014}, nano-silicates
\citep[][]{Hoang2016ApJ...824...18H, Hensley2017ApJ...836..179H}, or
spinning magnetic dipoles \citep[][]{HoangLazarian2016ApJ...821...91H,
  Hensley2017ApJ...836..179H}. Nano-silicates have, however, been
excluded as the sole source of the EME in the diffuse ISM
\citep[i.e. in {\em Planck} all-sky
maps,][]{Ysard2022A&A...663A..65Y}. A contribution to EME from the
(vibrational) thermal emission of magnetic dust
\citep[][]{DraineLazarian1999ApJ...512..740D} may be important in some
regions \citep[][]{DraineHensley2012ApJ...757..103D}.

Independent modelling efforts of spinning dust emission reach similar
predictions, for given dust parameters and local physical conditions
\citep[][]{Ali-Haimoud2009, Hoang2010ApJ...715.1462H,
  Ysard2010A&A...509A..12Y, Silsbee2011MNRAS.411.2750S,
  Zhang2025ascl.soft10015Z}. In addition, thermochemical PDR models
estimate the local physical conditions that result from the transport
of ultraviolet (UV) radiation \citep[][]{LePetit2006ApJS..164..506L}. Therefore,
observations of the cm-wavelength continuum in PDRs, and especially
its variations, can potentially calibrate the spinning dust models and
help identify the dust carriers. Eventually, the radio continua from
PDRs may provide constraints on physical conditions under the
spinning-dust hypothesis.

Cosmic Background Imager (CBI) observations have shown that the
surprisingly bright cm-wavelength continuum from the nearby
$\rho$\,Oph molecular complex peaks in the $\rho$\,Oph\,W PDR
\citep[][]{cas08, Arce-Tord2020MNRAS.495.3482A}. The {\em WMAP} and
{\em Planck} spectral energy distributions (SEDs) are fit by spinning
dust models \citep[][]{cas08,Planck2011A&A...536A..20P}. However, the
CBI maps revealed that the radio-IR correlation breaks down when
resolving the molecular complex. Under the spinning dust hypothesis,
this points at environmental factors that strongly increase the
spinning dust emissivity per nucleon ($j_{1\mathrm{cm}}/n_{\rm H}$) in
$\rho$\,Oph\,W, by a factor of at least 26 relative to the mid-IR (MIR) peak
in the whole complex \citep[][]{Arce-Tord2020MNRAS.495.3482A}.

Australia Telescope Compact Array (ATCA) observations of
$\rho$\,Oph\,W resolved the width of the filament
\citep[][]{Casassus2021MNRAS.502..589C}.  The multi-frequency
17--39\,GHz mosaics showed morphological variations in the
cm-wavelength continuum. The corresponding spectral variations
can be interpreted in terms of the spinning dust emission mechanism as
a minimum grain size cutoff at $6\pm1${\AA}, that increases deeper
into the PDR. This interpretation is strengthened by the qualitative
agreement with a PAH size proxy. It is also consistent with the
conclusions of studies of the Orion Bar, based on near- to far-IR (NIR-FIR) JWST
data, which take into account radiative transfer through PDRs
\citep[][]{Elyajouri2024A&A...685A..76E}.

The ATCA observations of $\rho$\,Oph\,W point at EME spectral
variations within a single source. Such spectral variations may be key
to calibrate spinning dust models and to identify the carriers and
their spin-up mechanism.  However, with only five antennas, the ATCA
39\,GHz data are noisy and strongly affected by flux loss due to
missing spatial frequencies (i.e. source angular size much larger than
the maximum recoverable scale). In addition the PDR peak at 39\,GHz is
contaminated by the SR\,4 point source, which is difficult to remove
at the angular resolution of the ATCA data (the width of the filament
is only a couple of beams).

The commissioning of the Band\,1 receivers on the Atacama Large
Millimetre Array (ALMA), covering 35\,GHz to 50\,GHz, offers the
opportunity to map the filament with much better fidelity than with
ATCA, and at finer angular resolutions. ALMA can also map the
cm-wavelength spectral index across the PDR, which, combined with the ATCA
17\,GHz and 20\,GHz data,  could allow for an exploration into the
spinning dust parameter space.

The same ALMA observations can be used to search for fine structure in
the EME spectrum, which may be used to identify its carriers
\citep[potentially through the detection of PAH
combs,][]{Ali-Haimoud2014, Ali-Haimoud2015MNRAS.447..315A}. The
Band\,1 data also allow to place limits on carbon recombination lines
(CRRLs). One of the spin-up mechanism is plasma drag (i.e. encounters
with H$^+$ and C$^+$ ions), which requires the EME to be coincident
with the C\,{\sc ii} region in the PDR.

Here, we report on mosaic observations of $\rho$\,Oph\,W in ALMA
Band\,1 at $\sim$7\,arcsec resolution (or twice as fine as the
39\,GHz ATCA data). Section\,\ref{sec:obs} summarises our observations
as well as our strategies for imaging and point-source
subtraction. Section\,\ref{sec:discussion} presents the extraction of the
Band\,1 spectra, a compilation of the multi-frequency SEDs, and a discussion of the limits
on any fine structure in the spectrum, both on PAH combs and CRRLs.
Section\,\ref{sec:conc} gives our conclusions. An interpretation of the measured
SEDs in terms of the spinning dust hypothesis will be presented in a
companion article.

\section{Observations} \label{sec:obs}

\subsection{The  $\rho$\,Oph\,W filament and the breakdown of the radio-IR correlation}

The region of the $\rho$\,Ophiuchi molecular cloud exposed to UV
radiation from HD\,147889 (i.e. $\rho$\,Oph\,W) is among the closest
examples of PDRs, located at a distance of 138.9\,pc. It is seen edge-on and
extends over $\sim$10$\times$3\,arcmin. It has been
extensively studied in the FIR atomic lines observed by ISO
\citep{1999A&A...344..342L,Habart2003}.

The $\rho$\,Oph complex is a region of intermediate-mass star
formation \citep{White2015MNRAS.447.1996W,Pattle2015MNRAS.450.1094P}.
It does not host a conspicuous H\,{\sc ii} region, in contrast to the
Orion Bar, another well-studied PDR, where the UV fields are $\sim$100
times stronger. While the HD\,147889 binary has the earliest spectral
types in the complex \citep[B2{\sc iv} and B3{\sc iv},][]{cas08}, the
region also hosts two other early-type stars: S\,1 \citep[also a
binary including a B4{\sc v}
primary,][]{LadaWilking1984ApJ...287..610L} and SR\,3 \citep[with
spectral type B6{\sc v},][]{Elias1978ApJ...224..453E}. Both S\,1 and
SR\,3 are embedded in the molecular cloud. Figure\,2 of
\citet[][]{Arce-Tord2020MNRAS.495.3482A} provides an overview of the
region, including the relative positions of these three early-type
stars.

Interestingly, the peak at all IR wavelengths, i.e. the circumstellar
nebula around S\,1, is undetectable in the CBI data. Upper limits on
the S\,1 flux density and correlation tests with {\em
  Spitzer}-IRAC\,8\,$\mu$m rule out a linear radio/IR relationship
within the CBI 45\,arcmin primary beam (which encompasses the bulk of 
the molecular cloud  mass). This breakdown of the radio-IR correlation in
the $\rho$\,Ophiuchi complex is further pronounced at finer angular
resolutions, with observations from the CBI\,2 upgrade to CBI
\citep[][]{Arce-Tord2020MNRAS.495.3482A}.

Thus, while the cm-wavelength and NIR-MIR signals in the
$\rho$\,Oph\,W filament display a tight correlation (as expected for EME), this
correlation breaks down in the $\rho$\,Oph complex as a whole, when
including the circumstellar nebula around S\,1. The brightest IR
PAH emission bands are observed near S\,1, but this region is undetectable at
31\,GHz.

The breakdown of the radio-IR correlation suggests that
cross-correlation techniques are of limited use in the interpretation
of the EME signal in $\rho$\,Oph.  Such cross-correlations have
successfully been applied to all-sky surveys, since the diffuse EME
signal can be described in terms of average physical conditions and
dust properties. However, in a PDR interface, the conditions are not
homogeneous.

\subsection{Data acquisition}

The ALMA Band\,1 observations of $\rho$\,Oph\,W were acquired in ALMA
program {\tt 2023.1.00265.S} using the 12\,m array. An observation log
is provided in Table\,\ref{table:log}. The correlator was tuned to four
spectral windows (spws), centred on 37.209\,GHz, 39.146\,GHz,
41.167\,GHz, and 43.104\,GHz, with 1920 channels each. The channel
width was set to 976.562\,kHz, providing a bandwidth of 1.875\,GHz per
spw. The data from the last execution block, on 14-Mar-2024, were
corrupted in several ways and were not included in the analysis.

The data were calibrated by staff from the North America ALMA Regional
Center, using CASA version 6.5.4.9 and ALMA pipeline version
2023.1.0.124. Full details on the  calibration can be obtained through the
ALMA science archive, in the quality assurance reports. In brief, some
flagging was applied, in particular to a channel at 43.91\,GHz and to
antenna DV05 for spw 39.146\,GHz. The phase and pointing calibrator
was J1625-2527, while the band pass and flux calibrator was J1617-5848.

\subsection{Imaging}

\begin{figure*}
  \centering
\includegraphics[width=\textwidth]{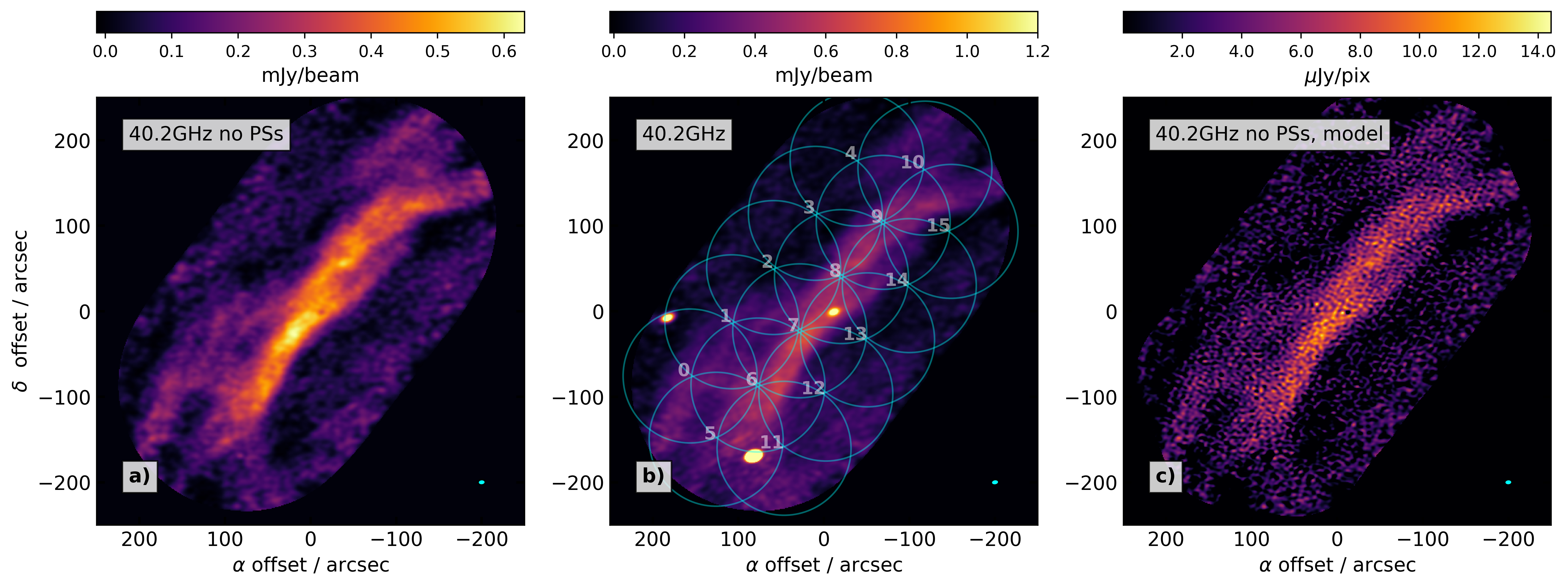}
\caption{ALMA Band\,1 continuum imaging of the
     $\rho$\,Oph\,W filament. Each image is shown in linear stretch,
     and $x-$ and
     $y-$ axis are offset along R.A. and Dec. from J2000 16:25:57.0
     $-$24:20:48.0. {\bf a)} Restored image, in natural weighting,
     after point-source subtraction, with a beam of
     8\farcs696$\times$6\farcs436/--86deg (see text), and a noise of
     5.55\,$\mu$Jy\,beam$^{-1}$.  {\bf b)} Same as panel a, but before
     point-source subtraction. The three bright point sources are the
     young stellar objects SR\,4 near the centre, ISO-Oph-17 to the
     east, and DoAr\,21 to the south. The cyan circles are centred on
     the phase directions of each field and their radii correspond to
     half the radius of the first Airy null (corresponding to a
     primary beam attenuation of
     $\sim$0.34). {\bf c)} Deconvolved model image,
     $I^m$, after  point-source subtraction and with
     1\arcsec$^2$ pixels.}
  \label{fig:mosaic}
\end{figure*}

The central region of the $\rho$\,Oph\,W filament was covered in a
mosaic of $K=16$ fields, as shown in Fig.\,\ref{fig:mosaic}b. Imaging
was performed as an instance of the maximum-entropy method, by fitting
a model image, covering $N=1024^2$ pixels, 1\arcsec$^2$ each, directly
to the visibility data. We used the {\tt gpu-uvmem} package
\citep{Casassus2006ApJ...639..951C, Carcamo2018A&C....22...16C} to
minimise the objective function,
\begin{equation}
  L = \sum_{k=1}^{K} \sum_{l=1}^{L_k} \omega_{k,l} \left| V_{k,l} - V^m_{k,l}\right|^2 - \lambda \sum_{i=1}^{N} p_i \ln(p_i), \label{eq:L}
\end{equation}
where $\omega_{k,l} = 1 / \sigma^2_{k,l}$ is the weight associated to
each visibility datum $V_{k,l}$, and $L_k$ is the total number of
visibilities for field $k$. The model visibilities $V^m_{k,l}$ are
computed on the model image $I^m(\vec{x}_i)$. The free parameters are
the intensities in each image pixel in units of a fiducial noise value,
$p_i = I^m_i / \sigma_I$, where $\sigma_I = 100$\,$\mu$Jy\,beam$^{-1}$
was chosen to be large as a pre-conditioning factor for the
optimisation. The theoretical noise of the natural-weights dirty map
is 1.53\,$\mu$Jy\,beam$^{-1}$, with a beam of
8\farcs70$\times$6\farcs44/--85.6 (in the format major axis $\times$
minor axis / position angle).

The model visibilities, $V^m$, are computed from the model image $I^m$
using the Airy-disk approximation to the ALMA primary beam
\citep[][]{ALMA_technical_handbook2019athb.rept.....R}, and by
propagating the model image to each frequency using a flat spectral
index in flux density ($\alpha = 0$). No visibility gridding was used.
Image restoration was carried out by producing a linear mosaic of the
residual dirty maps, in natural weights, and by adding the model image
smoothed by the Gaussian beam. The linear mosaic introduces an
attenuation map $\mathcal{A(\vec{x})}$ \citep[e.g. Eqs. A8 and A9
in][]{Casassus2021MNRAS.502..589C}.

The regularisation parameter $\lambda$ in Eq.\,\ref{eq:L} allows for
smoother model images than pure $\chi^2$ reconstructions, which tend
to fit the noise. Trials resulted in satisfactory results for
$\lambda = 0.1$, with thermal residuals. The corresponding images are
shown in Fig.\,\ref{fig:IRACsimul}. However, with entropy
regularisation, the spectral extraction technique turned out to be
significantly biased (see Appendix\,\ref{sec:IRACcalib}). We therefore
opted to set $\lambda=0$ and truncated the optimisation at ten
iterations to avoid excessive clumping.

The resulting model image, $I^m$, that fits the visibility data in a
pure least-squares sense, is shown in Fig.\,\ref{fig:mosaic} (but after
point-source subtraction; see Sect.\,\ref{sec:PSs}
below). We note that this image has not been convolved or smoothed in any
way. Its effective angular resolution is about one-third the natural-weight
clean beam \citep[e.g.][and references
therein]{Casassus2021MNRAS.507.3789C}.

Systematics in synthesis imaging are particularly strong when the
longest dimension of the source are resolved-out. In the case of the
ATCA observations, with five antennas and, hence, a very sparse
$uv$-coverage, we relied on the
IRAC\,8\,$\mu$m image as a prior to recover the missing interferometer
spacings. This choice was justified by the very tight
8\,$\mu$m-radio correlation at 17\,GHz and 20\,GHz. However, at higher
frequencies the radio filament is shifted parallel to its signal at 
IRAC\,8\,$\mu$m.  Still, with
$\sim$39 antennas and high-fidelity imaging, ALMA recovers the
$\rho$\,Oph\,W filament without the need of a prior, but with
significant flux loss. As described in Appendix\,\ref{sec:IRACcalib}, the flux recovered over the whole mosaic
is $\sim$30\%.

\begin{table*}
\caption{\label{table:log} Observation log.}
\centering
\begin{tabular}{lccccc}
\hline\hline
UT-Date      & $\Delta t$\tablefootmark{a}  & Elevation & pwv\tablefootmark{b}  & Phase rms  & Baseline \\
      &   (s)                                  & (deg)  & (mm)   & (deg) &range (m) \\
\hline
2024-03-11:11:17   &  4192    &  58 & 4.8 & 1.48 & 15-361 \\
2024-03-12:10:35   &  4131    &  67 & 6.5 & 1.40 & 15-314 \\
2024-03-13:11:06   &  4128    &  59 & 7.6 & 1.85 & 15-314 \\
2024-03-14:11:29   &  4242    &  53 & 8.2 & 6.12 & 15-314 \\
\hline
\end{tabular}
\tablefoot{The number of antennas ranged from 36 to 39. 
\tablefoottext{a}{Time on-source}
\tablefoottext{b}{Mean column of precipitable water vapour.}
}
\end{table*}

\subsection{Point-source subtraction} \label{sec:PSs}

There are three bright point sources in the mosaic of $\rho$\,Oph\,W,
corresponding to the young stellar objects SR\,4, ISO-Oph-17, and
DoAr\,21. These were subtracted in the $uv$ plane, handling the raw
(non-gridded) visibility data with the {\tt pyralysis} package
\citep[][]{casassus_carcamo2022MNRAS.513.5790C}. We first fit for the
point source positions, their flux densities, $F_{\nu_\circ}$, at the
reference frequency, $\nu_\circ = 40.15\,$GHz (corresponding to the
centre frequency of the four spectral windows), and spectral indices,
$\alpha$. In order to reduce the volume of the data, we averaged the
visibilities into 60\,channel bins (using {\tt CASA} task {\tt
  mstransform}). The point-source fits were carried out for each
execution block separately, and selecting the fields closest to each
point source.  We used the {\tt iminuit} package
\citep[][]{James:1975dr,iminuit}, using the Hessian approximation to
estimate uncertainties.  Provided with models for each point source,
we proceeded with their subtraction from the non-channel-averaged
visibilities.

The imaging of the point-source-subtracted dataset shown in
Fig.\,\ref{fig:mosaic}a offers satisfactory results. The point-source
fits to DoAr\,21 required the incorporation of the four nearest fields to reach
thermal residuals. We attribute the imperfect point-source subtraction
for DoAr\,21 in single-field point-source fits (with $\lesssim 5\%$
residuals) to the accuracy of the Airy disk approximation to the ALMA
primary beam.

\begin{figure}
  \centering
   \includegraphics[width=\columnwidth]{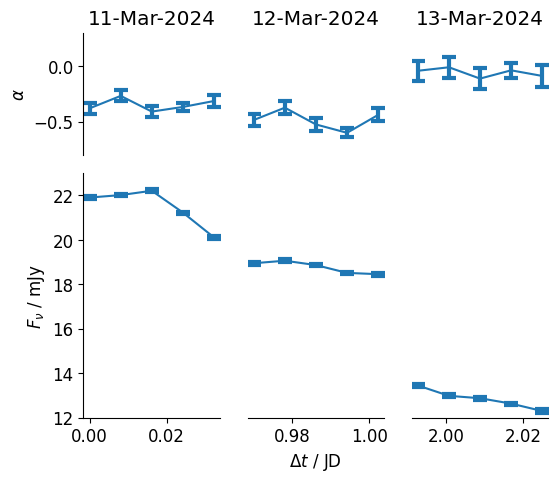}
   \caption{Point source fits to DoAr\,21, over individual interferometer scans, each $\sim$10\,min long. The top and bottom rows respectively record   spectral index, $\alpha$, and flux density, $F_\nu$, at the reference frequency of 40.15\,GHz. The $x-$axis gives the Julian date, expressed in days  from the start of the observations. }
  \label{fig:DoAr21}
\end{figure}

As summarised in Table\,\ref{table:PSs}, the flux densities for
ISO-Oph\,17 and SR\,4 were consistent in each execution block, with no
indication of variability. The spectral indices for these two sources
are consistent with the Rayleigh-Jeans tail of the sub-mm dust
continuum.  However, the very bright non-thermal emission from
DoAr\,21 turned out to be highly variable. This star is known to be a
compact binary. VLBA astrometric monitoring is consistent with two
sub-stellar companions in external orbits
\citep[][]{Curiel2019ApJ...884...13C}. DoAr\,21 has no detectable disk
in the sub-mm continuum \citep[][]{Cieza2019MNRAS.482..698C}. Our
measurements for DoAr\,21 are summarised in Fig.\,\ref{fig:DoAr21}.

\begin{figure}
  \centering
   \includegraphics[width=\columnwidth]{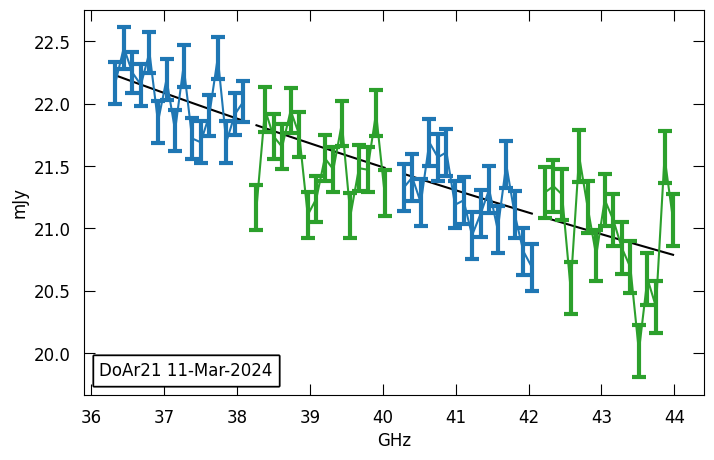}
   \caption{Flux densities for DoAr\,21, on 11-Mar-2024, extracted over  120-channel averages. The black line is the best fit point-source model from Table\,\ref{table:PSs}. Spectral windows are plotted in alternating colours.}
  \label{fig:DoAr21spec}
\end{figure}

\begin{figure}
  \centering
   \includegraphics[width=\columnwidth]{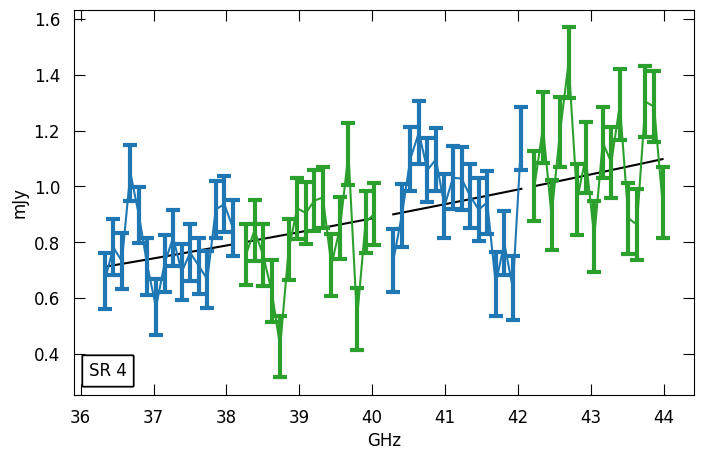}
   \caption{Flux densities for SR\,4, extracted over  120-channel averages. The black line is the best fit point-source model from Table\,\ref{table:PSs}. Spectral windows are plotted in alternating colours.}
  \label{fig:SR4spec}
\end{figure}

As a sanity check on the point-source fits (and provided with their
best-fit positions, $\vec{x_\circ}$), we also extracted point-source
flux densities in each channel. We selected the closest field to each
source, and obtained the point-source flux densities with
\begin{equation}
  F_\nu  = \frac{\sum_{l=1}^{L_k} \omega_{k,l}~ \Re\left[  V_{k,l} e^{-2\pi i \vec{u}_{k,l}\cdot\vec{x_\circ}}\right]  }{ \mathcal{A}_k(\vec{x_\circ})  \sum_{l=1}^{L_k} \omega_{k,l} }
,\end{equation}
where $\Re$ represents the real part, $\mathcal{A}_k$ is the
primary-beam attenuation pattern for field $k$, and $\vec{u}_{k,l}$ is
the $uv$ plane coordinate for datum $V_{k,l}$. The associated errors
are
\begin{equation}
  \sigma(F_\nu) = \frac{1}{ \mathcal{A}_k(\vec{x_\circ})  \sqrt{\sum_{l=1}^{L_k} \omega_{k,l} }  }. \label{eq:sigmaF}
\end{equation}
Example spectra, extracted over all epochs, are shown in
Figs.\,\ref{fig:DoAr21spec} and \,\ref{fig:SR4spec}, for DoAr\,21 and
SR\,4. The spectral index fits from Table\,\ref{table:PSs} are
consistent with the channel flux densities.

\begin{table*}
  \caption{\label{table:PSs} Point source flux densities 
    and spectral indices at 40.15\,GHz.}
\centering
\begin{tabular}{lccc}
\hline\hline
        & 2024-03-11    & 2024-03-12 & 2024-03-13 \\
\hline
  SR\,4        & ($860\pm28$)/($3.14\pm 0.56$) &  ($886\pm 27$)/($1.51\pm 0.55$)  & ($942\pm 30$)/($2.1\pm 0.58$)  \\ 
                & \multicolumn{3}{c}{($893\pm16$)/($2.27\pm0.32$)\tablefootmark{a}} \\ \hline
  ISO-Oph\,17  &  ($2507 \pm 49$)/($2.64\pm 0.36$)    &  ($2562\pm 48 $)/($3.61\pm 0.33)$  &   ($2720\pm 54$)/($2.92\pm 0.36$) \\ 
                & \multicolumn{3}{c}{($2588 \pm 29$)/($3.09\pm 0.20$)\tablefootmark{a}} \\ \hline
  DoAr\,21       &  ($21462\pm26$)/($-0.35\pm0.02$)    &  ($18731\pm 25$) / ($-0.502\pm 0.02$) &     ($12819\pm 28$) / ($-0.06\pm0.04$ )\\
\hline
\end{tabular}
\tablefoot{Flux densities 
    and spectral indices are given in the format $(F_\nu) / (\alpha)$, with
    $F_\nu$ in $\mu$Jy.  These
    measurements do not include a systematic $\sim$5\% flux
    calibration uncertainty. 
\tablefoottext{a}{The fits to all three execution blocks were
    carried out on the concatenated visibility data.}
}
\end{table*}

\subsection{Self-calibration trials}\label{sec:selfcal}

As explained in Appendix\,\ref{sec:cleanvsguvmen}, after point-source
subtraction {\tt gpu-uvmem} generates images with much higher dynamic
range than achievable with CASA-{\tt tclean}. However, {\tt gpu-uvmem}
does not perform as well for extended signal in the vicinity of bright
point sources. We therefore attempted self-calibration of the
60-channel average data, using the {\tt gpu-uvmem} model visibilities
for the extended signal, but with the addition of the point-source
best fits. Given the variability of DoAr\,21, this trial was attempted
in individual execution blocks. We performed a single iteration of
tasks {\tt gaincal} and {\tt applycal}, using phase-only gain
calibration, and without additional flagging. Imaging with {\tt
  gpu-uvmem} for the 11-Mar-2024 execution block, after  point-source
subtraction, yielded a peak signal of 733\,$\mu$Jy\,beam$^{-1}$ and a
noise level of 7.2\,$\mu$Jy\,beam$^{-1}$ (or a signal-to-noise ratio of S/N=101.81 (for a beam of $8\farcs912 \times 6\farcs050$). For
comparison, before self-calibration the same restoration yields
essentially the same value, with a somewhat larger S/N=101.96. We
concluded that self-calibration was not necessary for this dataset.

\section{Analysis and discussion} \label{sec:discussion}

\subsection{Limits on carbon recombination lines} \label{sec:CRRLs}

There are four carbon radio-recombination lines (CRRLs) within our
correlator setup, from C56$\alpha$ to C53$\alpha$. The turbulent
linewidths in the filament are likely of the order of $\sim$2\,km\,s$^{-1}$
\citep[][]{Pankonin1978A&A....64..333P}; thus, they are narrower than the channel
width of $\sim$7\,km\,s$^{-1}$. We stacked all four lines in a joint
imaging and obtained a limit of 0.15\,mJy\,beam$^{-1}$, with a
$8\farcs81 \times 6\farcs53$ beam.  The physical conditions in
$\rho$\,Oph\,W inferred from PDR models \citep[][]{Habart2003} should
yield LTE line intensities of $\sim$2\,mJy\,beam$^{-1}$ \citep[at
$\sim 300\,$K,][]{Casassus2021MNRAS.502..589C}. Depopulation factors
are negligible above 100\,K
\citep[][]{Salgado2017ApJ...837..141S}. The non-detection of CRRLs is
surprising and might be due to collisional de-excitation of the
denser regions in the filament, resulting in an unstructured map,
which might have beeen filtered out in the interferometer observations.

\subsection{Extraction of the intra-band diffuse emission spectrum} \label{sec:diffusespec}

The diffuse emission from $\rho$\,Oph\,W is too faint for the extraction of its spectrum in individual channels through aperture photometry. Instead, we opt for cross-correlation with a template image, for which we use the deconvolved model image, $I^m$, of Fig.\,\ref{fig:mosaic}c, assuming a flat spectral index. We search for 
a linear-regression slope, $a_\nu$, such that
\begin{equation}
I_\nu = a_\nu I^m,  
\end{equation}
by simulating visibilities on the template image, $V^m_{k,l}$   \citep[using the {\tt pyralysis} package,][]{casassus_carcamo2022MNRAS.513.5790C}, and minimising
\begin{equation}
  \chi^2 = \sum_{k=1}^{K} \sum_{l=1}^{M_k} \omega_{k,l} \left| V_{k,l} - a_\nu V^m_{k,l}\right|^2 ,
\end{equation}
where the sum is restricted to the $M_k < L_k$ visibilities at frequency $\nu$ and for field $k$. The result is
\begin{eqnarray}
  a_\nu & = & \frac{ \sum_{k=1}^{K} \sum_{l=1}^{M_k} \omega_{k,l}  \Re{\left[  V_{k,l}^{\ast} V^m_{k,l} \right] }  }{\sum_{k=1}^{K} \sum_{l=1}^{M_k} \omega_{k,l}  \left| V^m_{k,l}  \right|^2    }, \\
  \sigma(a_\nu) & = & \sqrt{\frac{ 1}{\sum_{k=1}^{K} \sum_{l=1}^{M_k} \omega_{k,l}  \left| V^m_{k,l}  \right|^2 }}.
\end{eqnarray}

The cross-correlation technique used here mitigates against the effect
of flux loss or missing spatial frequencies in images reconstructed
from interferometric data. A residual bias is still expected in the
determination of spectral slopes, since $uv-$coverage varies with
frequency. This bias is studied in Sec.\,\ref{sec:IRACcalib}. In
pure-$\chi^2$ imaging (i.e. without the introduction of entropy
regularisation as per  Eq.\,\ref{eq:L}), this residual bias in spectral
slope is negligible. It does, however, become detectable with the
introduction of entropy, which is the reason we avoided
regularisation.

\begin{figure}
  \centering
\includegraphics[width=\columnwidth]{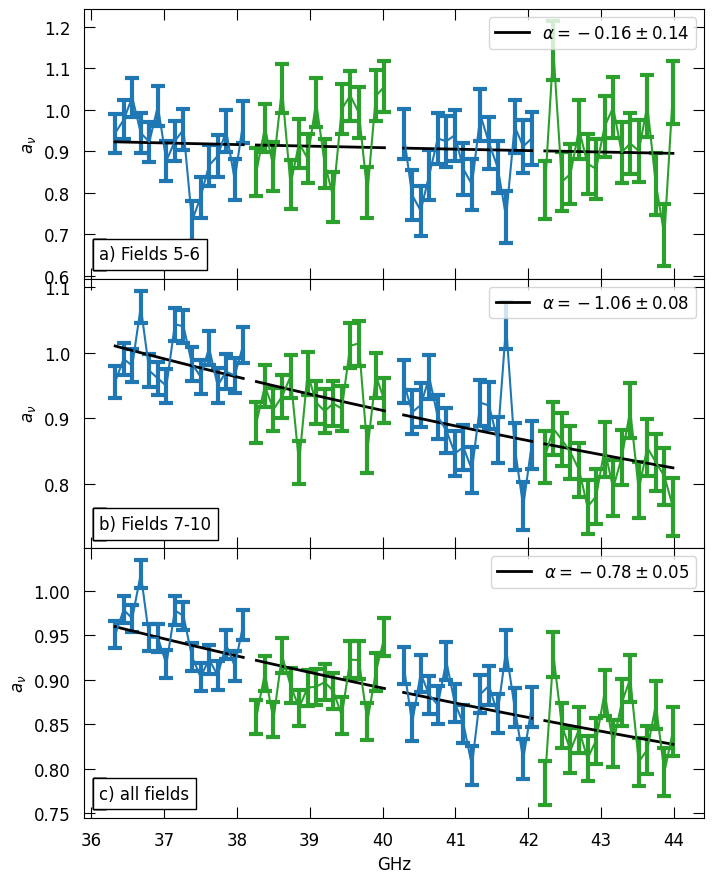}
\caption{Spectrum of the diffuse emission from $\rho$\,Oph\,W, as estimated with the cross-correlation slope, $a_\nu$. The black line is the best-fit power-law spectrum, with spectral index indicated in the legends. Plots a) to c) correspond to spectra extracted over different parts of the mosaic, as indicated in each plot.  Field IDs follow from Fig.\,\ref{fig:mosaic}. Spectral windows are plotted in alternating colours.}
  \label{fig:anu}
\end{figure}

Example spectra are shown in Fig.\,\ref{fig:anu} for the 120-channel
average and extracted over regions along the filament.  Power-law
fits for
\begin{equation}
  a_\nu \sim a_{\nu_\circ} (\nu/ \nu_\circ)^{\alpha_d},  \label{eq:diffusespecfit}
\end{equation}
yield varying spectral indices, $\alpha_d$, over different regions. The
spectral index variations are very significant along the filament,
from $\alpha_d = -0.16\pm0.14$ in fields 5 and 6 (south-east of the
filament), to $\alpha_d = -1.06\pm0.08$ in fields 7 to 10 (north-west
of the filament). The slopes $a_{\nu_\circ}$ are slightly biased
downwards, with $a_{\nu_\circ} = 0.96\pm0.005$ for
$\nu_\circ = 36.63\,$GHz averaged over all fields, which results from the
use of the pure-$\chi^2$ model image as template. In pure $\chi^2$
imaging, the model image fits the noise to some extent, which leads to
positive-definite noise that then acts as extra signal in the
simulated channel visibilities.  The same exercise using entropy
regularisation, with $\lambda = 0.1$, results in $a_\nu$ value that is consistent
with 1, but at the expense of a bias in $\alpha_d$ (see
Sect.\,\ref{sec:IRACcalib}).

\subsection{Cross-correlations with the {\em Spitzer}-IRAC maps} \label{sec:IRACxcorr}

As previously noted, the cm-wavelength continuum in $\rho$\,Oph\,W
is  closely correlated with the NIR emission, in particular as traced
in the four bands of the Infrared Array Camera (IRAC) aboard {\em
  Spitzer} \citep[as provided by the {\em c2d} Spitzer Legacy
Survey][]{Evans2009ApJS..181..321E}. The IRAC images were obtained
from the Set of Enhanced Imaging Products
(SEIP)\footnote{\url{https://doi.org/10.26131/IRSA433}}, which is part
of the Spitzer Heritage Archive. Point sources were removed prior to
using the IRAC maps as cross-correlation templates for radio
interferometric simulations.  Details on  IR point-source subtraction
are given in Appendix \,\ref{sec:IRACPSs}.

We quantified the degree of correlation between the ALMA Band\,1 data
and the IRAC bands using linear regression and the Pearson $r$
cross-correlation coefficient. IRAC angular resolutions are
1$''\!\!$.66, 1$''\!\!$.72, 1$''\!\!$.78, and 1$''\!\!$.98 at 3.6,
4.5, 5.6, and 8.0\,$\mu$m, respectively \citep[][]{Fazio2004}. Since
most of the 40.2\,GHz signal is in the shortest spacings, and the
angular resolution of the ALMA map is $\sim 7$\,arcsec, biases in the
correlations due to the coarser resolution at 8.0\,$\mu$m are
negligible.

As in Sect.\,\ref{sec:diffusespec},  the radio-IR slope, 
 $a_{\rm IR}$, in specific intensities, $I_{\rm Band\,1} = a_{\rm  IR} I_{\rm IRAC}$, is given by 
\begin{equation}
a_{\rm IR} =  \frac{\sum_{k=1}^{K} \sum_{l=1}^{L_k} \omega_{k,l} \Re\left[V^*_{k,l} V^{\rm IRAC}_{k,l}\right]}{\sum_{k=1}^{K} \sum_{l=1}^{L_k} \omega_{k,l} \left|V^{\rm IRAC}_{k,l}\right|^2},
\end{equation}
where $\left\{V^{\rm IRAC}_{k,l}\right\}$ are the visibilities extracted on a simulation of the Band\,1 $uv$-coverage on the (point-source-subtracted) IRAC maps, obtained with the  {\tt pyralysis} package. The associated $1\sigma$ uncertainty is given by error propagation,
\begin{equation}
  \sigma(a_{\rm IR}) =  \frac{1}{\sqrt{ \sum_{k=1}^{K} \sum_{l=1}^{L_k} \omega_{k,l} \left|V^{\rm IRAC}_{k,l}\right|^2}}.
\end{equation}
The radio-IR cross-correlation is
\begin{equation}
C = \frac{ \sum_{k=1}^{K} \sum_{l=1}^{L_k} \omega_{k,l} \left(  V_{k,l} - \left\langle V\right\rangle  \right)^*  \left(  V^{\rm IRAC}_{k,l} - \left\langle V^{\rm IRAC}  \right\rangle \right)}{ \sum_{k=1}^{K} \sum_{l=1}^{L_k} \omega_{k,l}  },
\end{equation}
where the mean observed visibility  is 
\begin{equation}
  \left\langle V\right\rangle = \frac{ \sum_{k=1}^{K} \sum_{l=1}^{L_k} \omega_{k,l} V_{k,l} }{ \sum_{k=1}^{K} \sum_{l=1}^{L_k} \omega_{k,l}  },
\end{equation}
and likewise for the simulated visibilities  $\left\langle V^{\rm IRAC} \right\rangle$. The standard deviation of the Band\,1 data is
\begin{equation}
  \sigma_V =    \sqrt{\frac{ \sum_{k=1}^{K} \sum_{l=1}^{L_k} \omega_{k,l} \left( V_{k,l}  - \left\langle V\right\rangle\right)^2 }{ \sum_{k=1}^{K} \sum_{l=1}^{L_k} \omega_{k,l}  }},
\end{equation}
 and likewise for $\sigma_{V^{\rm IRAC}}$. Finally, an expression for the Pearson coefficient adapted to the case of complex visibility data is
\begin{equation}
r = \frac{ C}{\sigma_V \sigma_{V^{\rm IRAC}}}.
\end{equation}
The uncertainty on $r$ vanishes when computed over large samples,
which limits its use in discriminating between templates. However,
when taking the $r$ values in $uv$-radius bins the uncertainties can
become important for the larger $uv$-radii, because of the relative
scarcity of long baselines in this Band\,1 dataset. We estimated the
uncertainties in $r$ by bootstrapping.

\begin{table}
\caption{\label{table:IRACxcorr} Radio-IR slopes for each IRAC band.}
\centering
\begin{tabular}{cc}
\hline\hline
  3.6\,$\mu$m    &   $3.435\times 10^{-2} \pm 1.5\times 10^{-4}$ \\
  4.5\,$\mu$m  &  $6.345\times 10^{-2} \pm 3.0\times 10^{-4}$ \\
  5.8\,$\mu$m & $4.463\times 10^{-3} \pm 2.1\times 10^{-5}$ \\
  8.0\,$\mu$m  & $1.515\times 10^{-3} \pm 7.7\times 10^{-6}$  \\
  \hline
\hline
\end{tabular}
\tablefoot{These slopes are calculated over the whole Band\,1 visibility data, covering all fields and spectral windows.}
\end{table}

\begin{figure}
  \centering
\includegraphics[width=\columnwidth]{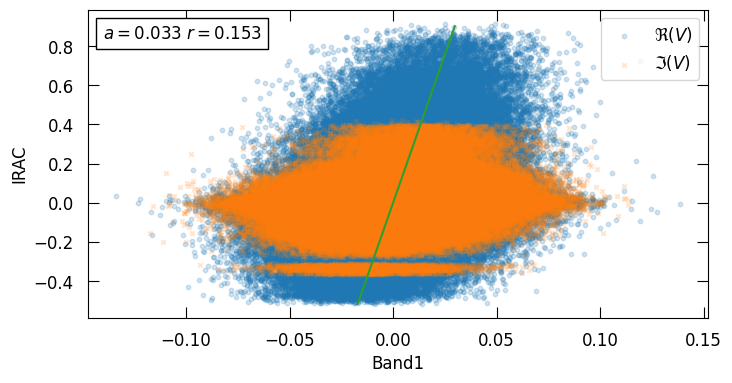}
\caption{\label{fig:IRACaslope} Radio-IR linear regression in real and imaginary parts, with Band\,1 visibilities on the $x-$axis, and simulated IRAC\,3.6\,$\mu$m visibilities on the $y-$axis. Axis units are given in Jy. Only data corresponding to $uv$-radii shorter than 2606$\lambda$ have been selected. The green line corresponds to the best fit linear regression. }
\end{figure}

Table\,\ref{table:IRACxcorr} lists the radio-IR slopes averaged over
all spectral windows and fields, and calculated after binning the data
into 16 channels (with a 120 channel average). The $r$ coefficients
turned out to be dominated by noise and essentially zero when using
all baselines. However, a cross-correlation signal is clear for short
baselines, containing the bulk of the signal, as illustrated in
Fig.\,\ref{fig:IRACaslope}.

\begin{figure}
  \centering
\includegraphics[width=\columnwidth]{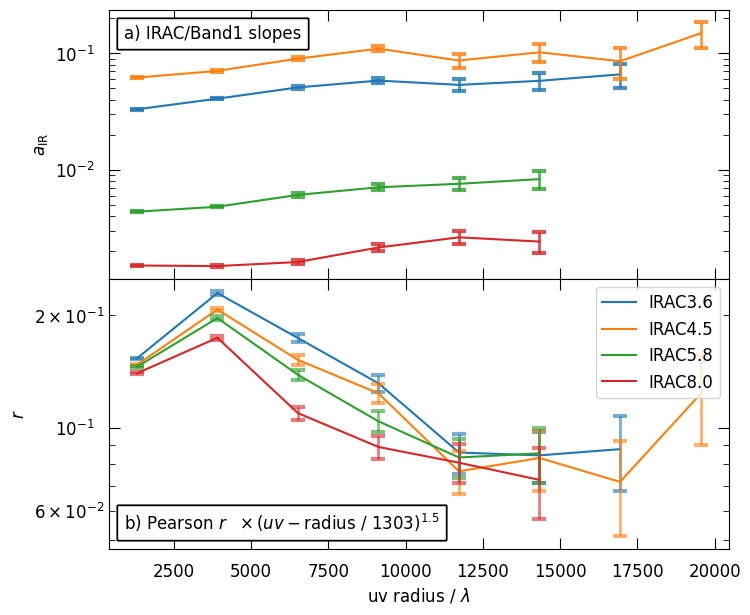}
\caption{\label{fig:IRACxcorr} Radio-IR slopes (in a) and Pearson $r$ coefficient (in b)  as a function of baseline length. The domain in $uv$-radius has been truncated at the maximum value for which the uncertainties become larger than a third of the reported values.}
\end{figure}

The trends in $a_{\rm IR}$ with $uv$-radius are summarised in
Fig.\,\ref{fig:IRACxcorr}a. Notably, the radio-IR slope increases with baseline length. For the best template
(i.e. 3.6\,$\mu$m), $a_{\rm IR}$ increases from
$3.3\times10^{-2}\pm 1.8\times 10^{-4}$ to
$5.83\times10^{-2}\pm 2.8\times 10^{-3}$ for the bin starting at
7818$\lambda$. This corresponds to a factor of $1.77\pm0.01$. Relative
to the IRAC power-spectrum, there is thus almost a factor of two more
power in Band\,1 at $\sim 20\arcsec$ than at $\sim 100\,\arcsec$. This
confirms that the Band\,1 and IRAC morphologies are appreciably
different, and limits the use of the IRAC bands as templates to
estimate the spectral index of the Band\,1 signal, since varying
frequencies correspond to varying $uv$-radii.

As shown in Fig.\,\ref{fig:IRACxcorr}b, the $r$ coefficients decrease
strongly with baseline, which we attribute to the radio-IR correlation
break-down when resolving the EME sources, even within the
$\rho$\,Oph\,W filament (see also Sect.\,\ref{sec:localpeak} below). In
addition, visibility measurements at longer baselines are less
frequent, and hence such $uv$-radii bins are noisier. In this case, the
$r$ values between each IRAC band are indistinguishable beyond
$\sim$10000$\lambda$.  The template with highest $r$ values is
IRAC\,3.6\,$\mu$m, and the ratio in $r$ between IRAC bands is greatest
for $uv$-radii between 5000 and 7000\,$\lambda$, corresponding to
spatial scales of $\sim 40\arcsec$ to $\sim 30\arcsec$.

\subsection{A compact EME source with no IR counterpart} \label{sec:localpeak}

\begin{figure*}
  \centering
\includegraphics[width=\textwidth]{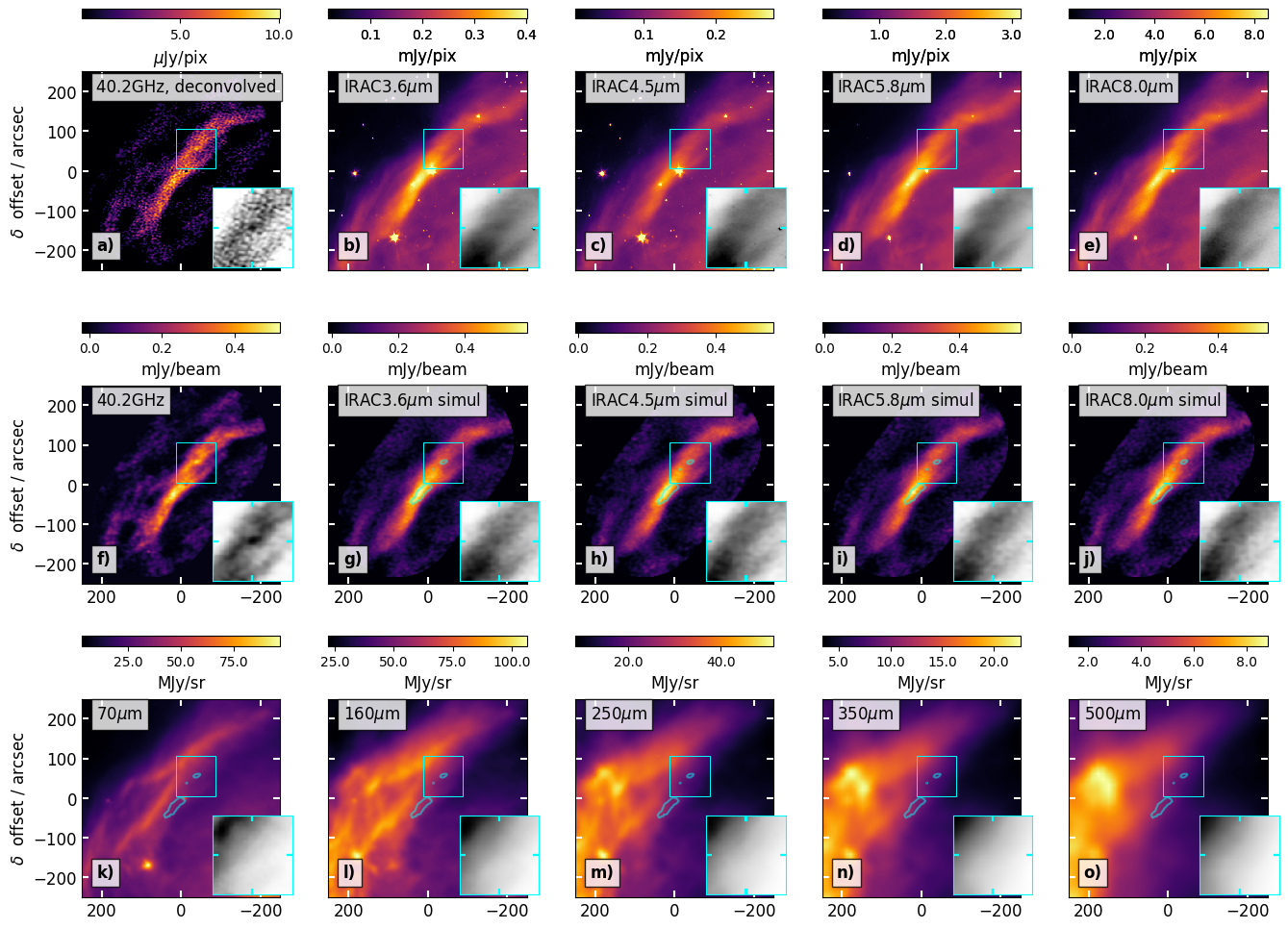}  
\caption{Comparison of the ALMA Band\,1 with IR templates. Insets highlight the compact EME source at  J2000 R.A. 16:25:54.154, Dec. -24:19:52.5.   The blue contour traces  the Band\,1 image from Fig.\,\ref{fig:mosaic}a  at 80\% peak.    a) Band\,1 deconvolved image (same as Fig.\,\ref{fig:mosaic}c, but with $\lambda=0.1$ for comparison). b-e) IRAC images,  with 1$\arcsec$ pixels, and before point-source subtraction. f) Restored Band\,1 image (same as Fig.\,\ref{fig:mosaic}a, but with $\lambda=0.1$). g-j)  Restorations of the IRAC simulations, after  point-source subtraction,  scaling by $a_{\rm IR}$, and  addition of thermal noise, as in the Band\,1 data. k-o) {\em Herschel} images.}
  \label{fig:IRACsimul}
\end{figure*}

As mentioned in Sect.\,\ref{sec:IRACxcorr}, the Band\,1 power at high
spatial frequencies increases relative to that of the IR signal. An
example in the image plane may be the strong peak found at J2000
right ascension (R.A.) of 16:25:54.154 and delination (Dec.) of $-$24:19:52.5, or about $\sim$45\arcsec
north-west from the centre of images in Fig.\,\ref{fig:mosaic}a (at an
offset of $\Delta \alpha = -38\arcsec$ along the R.A. and
$\Delta \delta = 55\arcsec$ along the Dec.). As shown in
Fig.\,\ref{fig:IRACsimul}a, this local peak appears to be point-like in
the Band\,1 model image, and reaches
$0.576\pm0.005\,$\,mJy\,beam$^{-1}$ in the restored image Fig.\,\ref{fig:IRACsimul}f, which rivals
the global peak along the filament, at
$0.631\pm0.005$\,mJy\,beam$^{-1}$.

The compact EME source is smaller than the $\sim 7\arcsec$ beam, and
bears no counterpart in the IRAC bands.  It is embedded along strong
and filamentary Band\,1 signal, which bends by $\sim 60$\,deg from the PDR
plane. Interestingly, this thin filament is absent from the IR maps,
which instead appear smooth and fainter relative to the peak. The
Band\,1 point source is not an artefact of the imaging strategy, as
verified through simulations of the Band\,1 $uv-$coverage on the IRAC
images (Fig.\,\ref{fig:IRACsimul}g-j; see also
Appendix\,\ref{sec:IRACcalib}). The point source has no FIR
counterpart either (Fig.\,\ref{fig:IRACsimul}k-o), as shown by
comparison with images from the {\em Herschel} Gould Belt Survey
\citep[][]{Herschel_2020A&A...638A..74L}.

\subsection{Comparison with the ATCA mosaics}

\begin{figure}
  \centering
\includegraphics[width=\columnwidth]{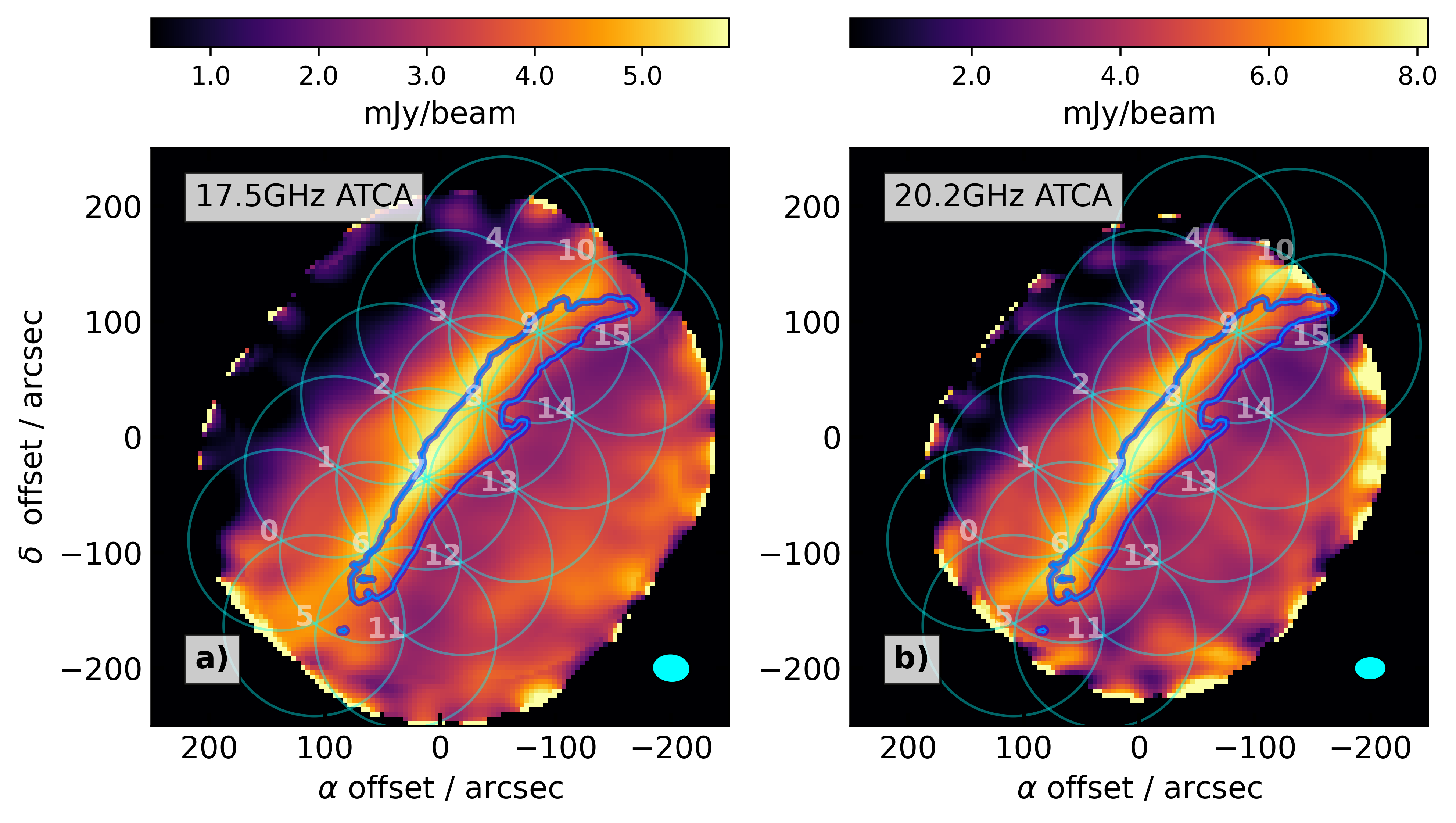}
\caption{\label{fig:ATCA} Comparison of the 17.5\,GHz (in a) and 20.2\,GHz (in b) ATCA mosaics \citep[from][]{Casassus2021MNRAS.502..589C} with the ALMA Band\,1 pointings. The Band\,1 restored image from Fig.\,\ref{fig:mosaic}a is overlaid in a single contour at 60\% peak intensity.}
\end{figure}

In the ATCA observations \citep[][]{Casassus2021MNRAS.502..589C}, at
30\,arcsec resolutions, the 17--20\,GHz intensities tightly follow the
MIR, i.e.  $I_{\rm cm} \propto I(8\,\mu{\rm m})$, despite the
breakdown of this correlation on larger scales. However, while the
33--39 GHz filament is parallel to IRAC\,8\,$\mu$m, it is offset by
15--20\,arcsec towards the UV source. This effect is also picked-up in
the Band\,1 40.2\,GHz observations, as shown in Fig.\,\ref{fig:ATCA}.

The blue-ward shift of the EME spectrum on the edge of the PDR closer
to HD\,147889 can be related to a gradient in physical conditions or
grain properties.  In particular, the trends in the ATCA data were
reproduced in the spinning dust hypothesis, in terms of a gradient in
minimum grain size, increasing deeper in the cloud
\citep[][]{Casassus2021MNRAS.502..589C}.

The present ALMA data cannot, however, be used to constrain resolved
gradients in physical conditions across the PDR. As explained below,
the Band\,1 observations must be corrected for flux loss to build a
multi-frequency SED, which can only be achieved, to a reasonable
accuracy, on scales corresponding to the primary beam. The Band\,1
spectral index measurements are also extracted over such large
scales. The combination of 7\,m (ALMA Compact Array) observations with
the present 12\,m observations could allow for such a resolved
multi-frequency analysis.

\subsection{Spectral energy distribution} \label{sec:SED}

Given the interferometer filtering, it is difficult to extract SEDs on
small scales; namely, of a single synthesised clean beam or
$\sim 7\arcsec$. Instead, we opt to build SEDs on scales of a primary
beam, which also allows estimates of the Band\,1 spectral index. The
total Band\,1 flux densities are estimated via aperture photometry,
and by correcting for the flux loss due to missing interferometer
spacings, for which we use the simulations on the IRAC\,3.6\,$\mu$m
template (see Sect.\,\ref{sec:IRACcalib} in Appendix). This correction
depends on the chosen photometric aperture. The thermal errors are
negligible compared to absolute flux calibration errors, and we opt to
consider a fixed 10\% error on the flux densities extracted for the
radio data.

The IR and FIR SEDs were built from the IRAC and {\em Herschel}
images. The flux densities from the IR ancillary data were extracted
by aperture photometry; namely, by integrating the product of the IR
images with the linear mosaic attenuation map (see
Sect.\,\ref{sec:IRACcalib} in Appendix).

The ATCA images reported in \cite{Casassus2021MNRAS.502..589C} already
fill in the missing spatial frequencies by incorporation of prior
images in the synthesis imaging strategy. We therefore proceeded
similarly as for the IR ancillary data, with the choice of smaller
apertures to accommodate the ATCA mosaics (see
Fig.\,\ref{fig:ATCA}). The resulting SEDs are tabulated in
Table\,\ref{table:SED}. Plots for these SEDs and model comparisons
will be presented in a companion article.

\begin{table*}
\caption{Broad-band SEDs for selected regions, defined by the Band\,1 fields they comprise.    \label{table:SED} }
\centering
\begin{tabular}{cccccccccccccccc}
  \hline\hline
         fields &    $\Omega$\tablefootmark{a} &    $f_{\rm B1}$\tablefootmark{b}  &    3.6\tablefootmark{c} &    4.5\tablefootmark{c} &    5.8\tablefootmark{c} &    8.0\tablefootmark{c}   & 70\tablefootmark{c} &   160\tablefootmark{c} &   250\tablefootmark{c} &   350\tablefootmark{c}  &  500\tablefootmark{c} &    40.2\tablefootmark{d} &        $\alpha$\tablefootmark{e} &    20.2\tablefootmark{d} &    17.5\tablefootmark{d} \\
   &   &    &             Jy   &       Jy &         Jy &         Jy &         Jy &         Jy &         Jy &         Jy &   Jy &        mJy &  &                      mJy        & mJy \\
            all      &37.6       &0.29      & 19.3   &    11.8  &      160   &     463    &   4167  &     7548   &    3216  &     1380  &      540  &      240 &$-0.78\pm0.05$ &      -   &    -\\
           5,6      & 9.1       &0.27      &  5.5   &     3.3  &       45   &     128    &   1173  &     2218   &     975  &      426  &      169  &       64 &$-0.16\pm0.14$ &      -   &    -\\
      7-10      &15.9       &0.37      &  9.4   &     5.6  &       79   &     224    &   1858  &     3014   &    1204  &      501  &      193  &      144 &$-1.06\pm0.08$ &    -     &    -\\
             6      & 6.1       &0.33      &  3.9   &     2.3  &       32   &      89    &    801  &     1447   &     625  &      270  &      107  &       55 &$-0.24\pm0.17$ &    130   &   72\\
           7-9      &12.5       &0.40      &  7.7   &     4.6  &       64   &     182    &   1525  &     2495   &     1009  &      423  &      164  &      125 &$-1.05\pm0.08$ &    265   &  146\\
  \hline
\end{tabular}
\tablefoot{The quoted 40.2\,GHz flux densities must be divided by $f_{\rm B1}$ for comparison with the other frequency points. Uncertainties in the radio flux densities are dominated by the absolute calibration error, of $\sim$10\% at 1$\sigma$. 
\tablefoottext{a}{Flux extraction aperture in arcmin$^2$.}
\tablefoottext{b}{Fraction of flux density recovered in the quoted Band\,1 measurement.}
\tablefoottext{c}{Centers of IRAC and {\em Herschel} bands in $\mu$m.}
\tablefoottext{d}{Center frequencies  of ALMA and ATCA measurements in GHz.}
\tablefoottext{e}{Spectral index at 40.2\,GHz.}
}
\end{table*}

\subsection{Limits on PAH combs}

The new ALMA observations can also be used to search for
fine structure in the spinning dust spectrum. If due to
quasi-symmetric PAHs, the EME may be structured with evenly spaced
combs of rotational lines, separated by $\Delta \nu \sim$1\,MHz
\citep[][]{Ali-Haimoud2014, Ali-Haimoud2015MNRAS.447..315A}. The exact
spacing, $\Delta \nu$, can be derived from the PAH rotational constant and
inversely proportional to the grain moment of inertia, which is itself
a measure of the number of carbon atoms in the PAHs. Such comb-like
fine structures in the EME spectrum would confirm the spinning dust
hypothesis and would identify the carriers \citep[as performed in
TMC-1 for small PAHs,][]{McGuire2018Sci...359..202M,
  McGuire2021Sci...371.1265M}.

As shown by \citet{Ali-Haimoud2014}, the rotational spectrum of imperfect planar PAHs, with either D$_6$h or D$_3$h quasi-symmetry and a permanent dipole arising from defects\footnote{ such as super- or de-hydrogenation, or deuterium,  $^{13}$C, or N substitutions},  should appear as a regularly spaced frequency comb,
\begin{equation}
\nu_J = \Delta \nu_{\rm line} \left( J + \frac{1}{2} \right),
\end{equation}
with 
\begin{equation}
\Delta \nu_{\rm line} \approx 2 A_3 = 2 \frac{h}{8 \pi^2 I_3},
\end{equation}
to first order in defects and where $I_3$ is the largest moment of inertia. This principal moment of inertia can be related to the number of carbon atoms, $N_{\rm C}$, in the PAH \citep{Ali-Haimoud2014}, via
\begin{equation}
I_3 \approx   1.5\times 10^4 \left( \frac{N_{\rm C}}{54} \right)^2  {\rm amu}\,\AA^2.  
\end{equation}

We use the cross-correlation spectrum, $a_\nu$, to place limits on such PAH combs in an adaptation of the matched-filter technique proposed in 
\citet{Ali-Haimoud2014} and \citet{Ali-Haimoud2015MNRAS.447..315A}. A function $f(\nu)$ is decomposed into a basis of comb-like functions, $c_i$, and its projection onto a particular comb is
\begin{equation}
  \mathcal{S}_i = \frac{\langle f(\nu) \cdot  c_i(\nu) \rangle}{ \| c_i(\nu)  \|^2}, 
\end{equation}
where the dot product of two functions is  
\begin{equation}
\langle   f(\nu) \cdot g(\nu) \rangle = \sum_k \omega_k f(\nu_k) g(\nu_k), 
\end{equation}
and $\omega_k = 1 / \sigma_k^2$. 

In this  application, the domain of the comb basis is split into $S=4$ spectral windows, each composed of  $C=1920$ regularly-spaced channels and with channel widths. $\delta {\nu_s}$. The comb,  $c_{\nu_s}$, for a spacing, $\Delta \nu_{\rm line}$, in spectral window, $s$, is  generated with the following algorithm,
\begin{equation}
  J_s = \rceil\left( \nu_s/ \Delta \nu_{\rm line}  -\frac{1}{2}\right),
\end{equation}
where $\rceil$ is the nearest integer,
\begin{equation}
  \Delta_{J_s} \nu  = \nu_s  - \Delta \nu_{\rm line} \times  \left(J_s + \frac{1}{2} \right),
\end{equation}
define $\tilde{c}_{\nu_s} = 1$ if
\begin{equation}
| \Delta_{J_s} \nu |   <  \frac{\delta_{\nu_s}}{2},
\end{equation}
and $\tilde{c}_{\nu_s} = 0$ otherwise.
The comb is given by
\begin{equation}
 c_{\nu_{s,c}} = \tilde{c}_{\nu_{s,c}}  - \frac{\sum_{c=1}^{C} \omega_{\nu_{s,c}} \tilde{c}_{\nu_{s}}   }{ \sum_{c=1}^{C} \omega_{\nu_{s,c}} }.
\end{equation}

We first  rectify the signal  $a_\nu$  by division with the power-law fit of Eq.\,\ref{eq:diffusespecfit},
\begin{equation}
  b_\nu = \frac{a_\nu }{ a_{\nu_\circ} (\nu/ \nu_\circ)^{\alpha_d}},
\end{equation}
and correct for zero mean, 
\begin{equation}
  \delta b(\nu_{s,c}) = b(\nu_{s,c}) - \sum_{c^\prime=1}^{C} \omega_{s,c^\prime} b(\nu_{s,c^\prime}). 
\end{equation}
The dot product with the comb basis is 
\begin{equation}
  \langle \delta\,b(\nu) \cdot  c_i(\nu) \rangle = \sum_{s=1}^{S} \sum_{c=1}^{C} \omega_{s,c} \delta b(\nu_{s,c}) c_i(\nu_{s,c}).
\end{equation}
In summary, the projection of the rectified signal,   $\delta b(\nu_{s,c}),$ on  comb $c_j$ with  line spacing $\Delta\nu_{\rm line}$, is
\begin{equation}
  \mathcal{S}(\delta b) =   \frac{\langle \delta\,b(\nu) \cdot  c_i(\nu) \rangle}{\langle c_i(\nu) \cdot  c_i(\nu) \rangle},
\end{equation}
with an uncertainty of
\begin{equation}
  \sigma(\mathcal{S}(\delta b)) =   \sqrt{\frac{1}{\langle c_i(\nu) \cdot  c_i(\nu) \rangle}}.
\end{equation}

\begin{figure}
  \centering
\includegraphics[width=\columnwidth]{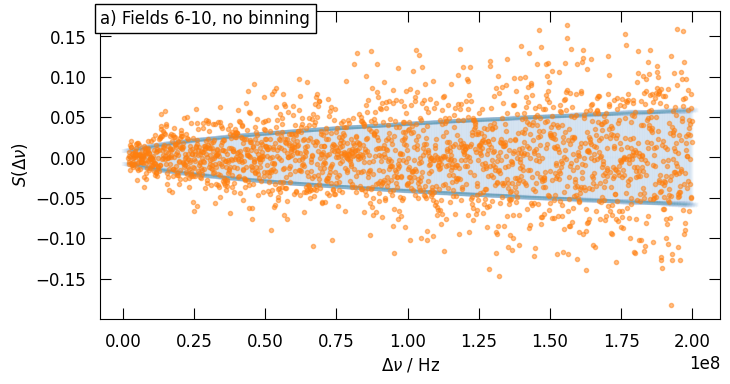}
\includegraphics[width=\columnwidth]{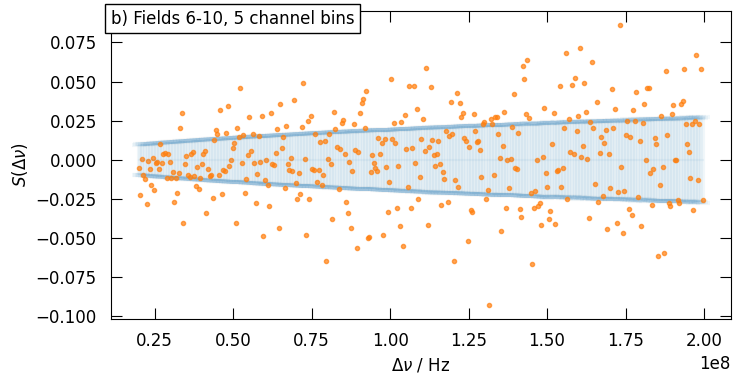}
\caption{Power $\mathcal{S}(\delta b)$ across a continuous range of  combs with line intervals $\Delta \nu$.  The top panel is for the raw spectral resolution, and the lower panel uses a five-channel average. Uncertainties are shown as vertical $\pm1\,\sigma$ error bars in blue (with total length of 2$\sigma$), and are offset to zero for  clarity. 
  \label{fig:Scombs}}
\end{figure}

The significance,    $\mathcal{S}(\delta b)/\sigma(\mathcal{S}(\delta b))$, is a measure of  the power on a comb basis, defined by the line interval $\Delta\nu_{\rm line}$. We scan $\Delta\nu_{\rm line}$ over the interval from twice the frequency resolution of $a_\nu$ (i.e. the  channel width, or its binned versions), and up to 200\,MHz, sampling every tenth of a resolution element. Fig.\ref{fig:Scombs} illustrates the resulting values for the case of five-channel bins and no channel binning.

\begin{figure}
  \centering
\includegraphics[width=\columnwidth]{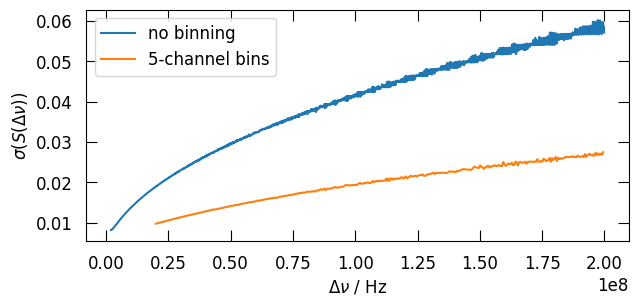}
\caption{Uncertainty on the projected  power in a continuous range of  combs with line intervals $\Delta \nu$. 
  \label{fig:Scombslimits}}
\end{figure}

There is no significant power in comb functions, with a spectral
resolution as fine as twice the channel width:
$\mathcal{S}(\delta b)/\sigma(\mathcal{S}(\delta b)) < 3.5$ in
$\sim$2000 values for $\Delta \nu_{\rm line}$. Since the signal has
been rectified and normalised, the uncertainty
$\sigma(\mathcal{S}(\Delta \nu_{\rm line}))$ directly gives the limit
of fraction of signal in combs with spacing $\Delta \nu_{\rm line}$,
as summarised in Fig.\,\ref{fig:Scombslimits}. For example, in native
spectral resolution, less than $\sim 3\times 0.06$ of the observed
flux density from $\rho$\,Oph\,W is due to combs with spacing of
$\sim$200\,MHz.

The absence of detectable fine-structure in the observed Band\,1
spectrum may reflect a large number of possible configurations for a
given number of carbon atoms
\citep[][]{Tielens2008ARA&A..46..289T}. In addition, the most stable
sub-nanometric particles are, a priori, neither flat nor symmetrical
\citep[][]{Parneix2017MolAs...7....9P,Bonnin2019PhRvA..99d2504B}. Furthermore, if the dipole moment is
boosted by atom substitution or gain (or loss) of H atoms, each possible
site of implantation would modify the moment of inertia and, thus, the
comb. All these effects lead to the conclusion that there are so many
weak lines superimposed that spectral structure cannot be detected, at
least within the present limits (Fig.\,\ref{fig:Scombslimits}).

\section{Conclusions} \label{sec:conc}

The ALMA observations of $\rho$\,Oph\,W, along with non-parametric
imaging and visibility-plane correlations, allow estimates of the
Band\,1 flux density and spectral index of the EME in
$\rho$\,Oph\,W. We place limits on the contribution from regularly
spaced spectral combs and CRRLs.  These spectral measurements will be
analysed in terms of the spinning dust hypothesis in a companion
article.

The Band\,1 image of $\rho$\,Oph\,W reveals bends and clumps along the
otherwise smooth IR filament. We compare the EME visibility power
spectrum against that of IR templates. Interestingly, for the best
template, i.e. IRAC\,3.6\,$\mu$m, there is almost a factor of two more
power in Band\,1 at $\sim 20\arcsec$ than at $\sim 100\,\arcsec$. An
extreme of such radio-only structures is a compact source of pure EME,
without any IR counterpart. It is embedded along strong and
filamentary Band\,1 signal, while the IRAC maps are featureless in the
same region.

\begin{acknowledgements}

  We thank the referee for constructive comments.  S.C. and
  M.C. acknowledge support from Agencia Nacional de Investigaci\'on y
  Desarrollo de Chile (ANID) given by FONDECYT Regular grant 1251456,
  and ANID project Data Observatory Foundation DO210001. This research
  has made use of data from the Herschel Gould Belt survey (HGBS)
  project (http://gouldbelt-herschel.cea.fr). The HGBS is a Herschel
  Key Programme jointly carried out by SPIRE Specialist Astronomy
  Group 3 (SAG 3), scientists of several institutes in the PACS
  Consortium (CEA Saclay, INAF-IFSI Rome and INAF-Arcetri, KU Leuven,
  MPIA Heidelberg), and scientists of the Herschel Science Center
  (HSC).
\end{acknowledgements}

\bibliographystyle{aa} \bibliography{refs.bib} 

\begin{appendix} 

  \section{Band\,1 flux loss and IRAC simulations} \label{sec:IRACcalib}

  Provided with the radio-IR slopes from Table\,\ref{table:IRACxcorr},
  we scale the IRAC visibilities simulated on the Band\,1 $uv$ plane,
  and add thermal noise as given by the observed visibility
  weights. We proceed to image the simulated IRAC data as for the
  Band\,1 observations. The resulting images are shown in
  Fig.\,\ref{fig:IRACsimul}.  It is 
  interesting that the peak signal in the 3.6\,$\mu$m image,
  found to yield the best Pearson $r$ coefficient with Band\,1, is
  indeed aligned with the Band\,1 peak, while the other bands shift
  farther from HD\,147889 (i.e. are shifted to the north-east) with
  increasing wavelength.

\begin{figure}
  \centering
\includegraphics[width=\columnwidth]{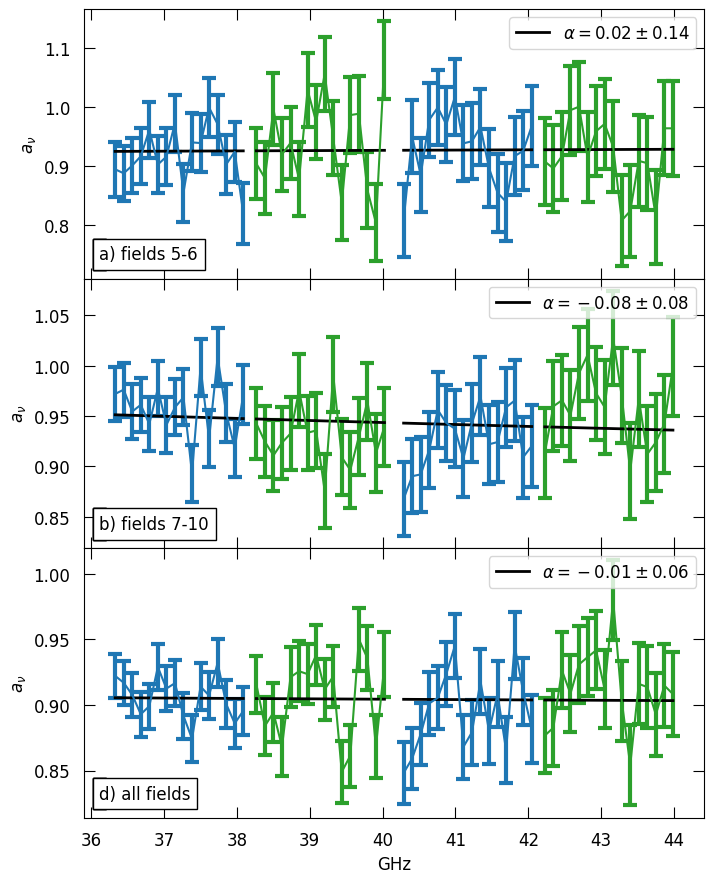}
\caption{Same as Fig.\,\ref{fig:anu}, but for the {\em IRAC}\,3.6\,$\mu$m simulation.}
  \label{fig:anu_IRAC}
\end{figure}

  The spectra shown in Fig.\,\ref{fig:anu_IRAC} follow from
  Fig.\,\ref{fig:anu}, but are extracted from the IRAC\,3.6\,$\mu$m
  simulation. We recover the input spectral index ($\alpha = 0$)
  within 1$\sigma$ in all cases. In contrast, the entropy-regularised
  images (with $\lambda=0.1$) yield biased spectral indices, for
  example $\alpha = -0.26\pm0.09$ for fields 7-10.

The IRAC\,3.6\,$\mu$m simulations allow estimates of the missing flux
due to the absence of short spacings in the central region of the
Band\,1 $uv$-coverage. The integrated flux density in the scaled
3.6\,$\mu$m template is 0.66\,Jy over the whole mosaic, as estimated
by integration of the product with the mosaic attenuation map
\citep[e.g. see the formulae given in the appendix of
][]{Casassus2021MNRAS.502..589C}. The same exercise on the restored
map yields 0.19\,Jy, so ALMA Band\,1 recovers about 29\% of the total
flux density.

The underlying assumption in our strategy for flux loss estimates is
that the Band\,1 signal follows the same power spectrum as the
template, which appears to hold on large scales, but deviates in small
scales (see Sect.\,\ref{sec:IRACxcorr}). However, the power spectra
drop sharply with increasing spatial frequency, so the effect of such
small scale deviations should be negligible on flux loss estimates on
large scales, such as in the Band\,1 primary beam.

\section{Clean versus gpu-uvmem comparison} \label{sec:cleanvsguvmen}

In order to compare the image restoration obtained with {\tt
  gpu-uvmem} against a standard CLEAN-based approach, we imaged the
point-source subtracted and channel-averaged visibilities using the
{\tt CASA-tclean} package \citep{CASA2022}. We used the {\tt mtmfs}
deconvolver, assuming a flat spectrum for the emission ({\tt
  nterms=1}), and adopted the Briggs weighting scheme with {\tt
  robust=2} (natural weighting) to maximise sensitivity to
diffuse emission.

Multi-scale CLEAN better accounts for extended structure. We employed
a set of angular scales ranging from point-like emission to several
times the synthesised beam (via the parameter {\tt
  scales=[0,15,38,70]}, specified in units of image pixels, with a
pixel size of $1''$). In addition, a low loop gain ({\tt gain=0.05})
was adopted to ensure stable convergence and to mitigate CLEAN bias
when deconvolving low-surface-brightness emission.

\begin{figure}
  \centering
   \includegraphics[width=\columnwidth]{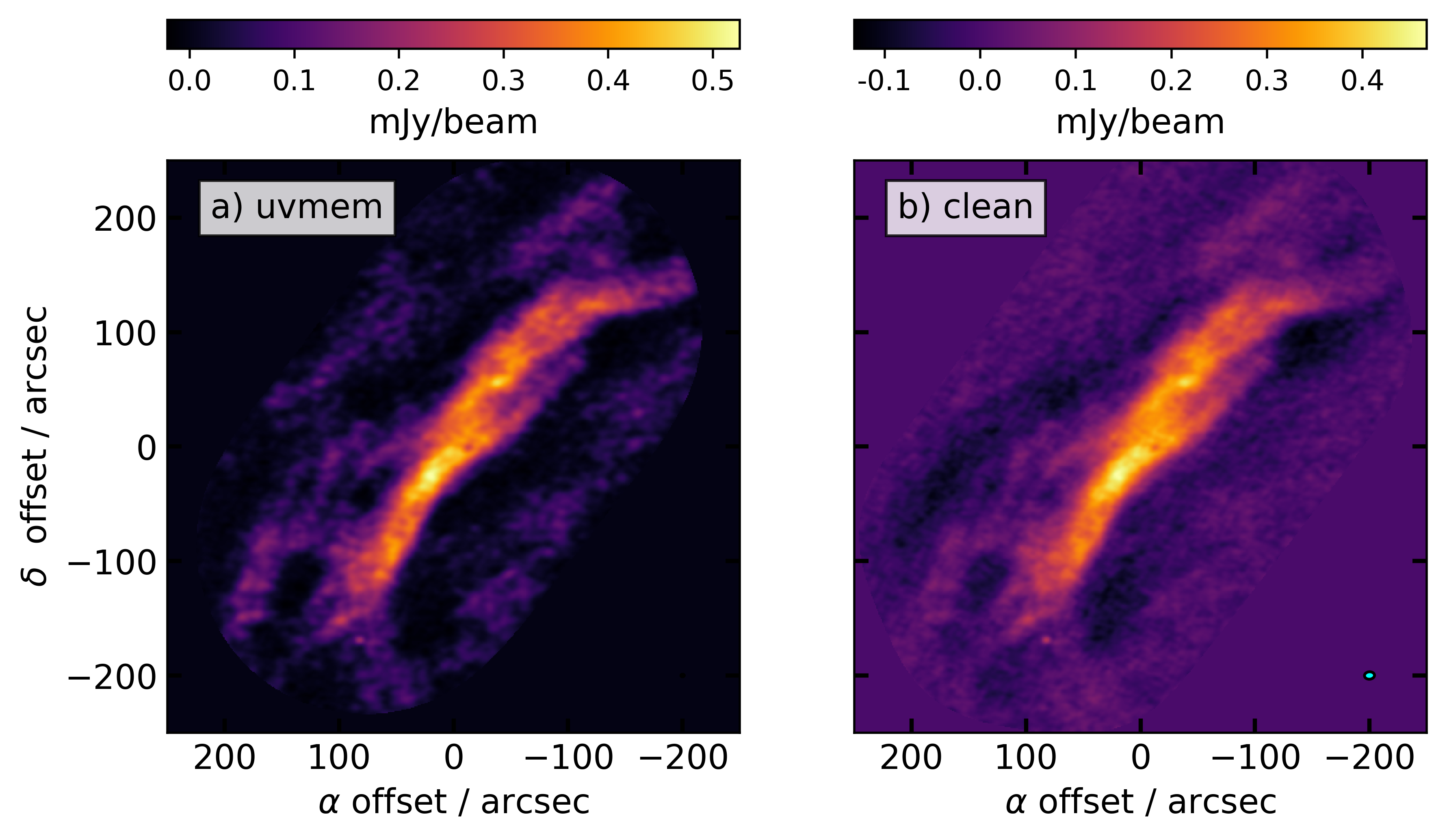}
   \caption{Comparison between {\tt gpu-uvmem} (in a), same as Fig.\,\ref{fig:mosaic}a) and {\tt tclean} reconstructions (in b).}
  \label{fig:clean}
\end{figure}

Fig.\,\ref{fig:clean} compares the {\tt gpu-uvmem} reconstruction with
the corresponding {\tt CASA-tclean} image. In the main region of
emission, the two reconstructions are very similar.  However, on the
largest angular scales the CLEAN image exhibits negative bowls,
reflecting the limited ability of CLEAN to recover smoothly
distributed emission in the presence of incomplete short-spacing
coverage. In contrast, {\tt gpu-uvmem} naturally suppresses
large-scale negative artefacts, leading to a more physically plausible
reconstruction in low-signal regions.

The synthesised beam size of the {\tt CASA-tclean} image is
$8.75''\times6.47/-85.49''$ (in the format major axis $\times$ minor
axis / position angle). The root-mean-square (rms) noise is
11\,$\mu$Jy\,beam$^{-1}$, and the signal-to-noise ratio of the peak of
the map is 42.

\section{Point-source subtraction in the IRAC images} \label{sec:IRACPSs}
  
The point sources in the IRAC images are predominantly normal stars, whose
MIR emission arises mainly from the Rayleigh-Jeans tail of their
stellar spectra. Such stars are expected to contribute negligibly at
radio wavelengths. Retaining stellar point sources would introduce
structures with no radio counterpart, biasing simulated visibilities
and artificially enhancing small-scale power. Their removal ensures
that the template represents only the diffuse emission relevant for
comparison with radio observations and for estimating the
interferometric flux recovery.

\begin{figure*}
  \centering
   \includegraphics[width=0.7\textwidth]{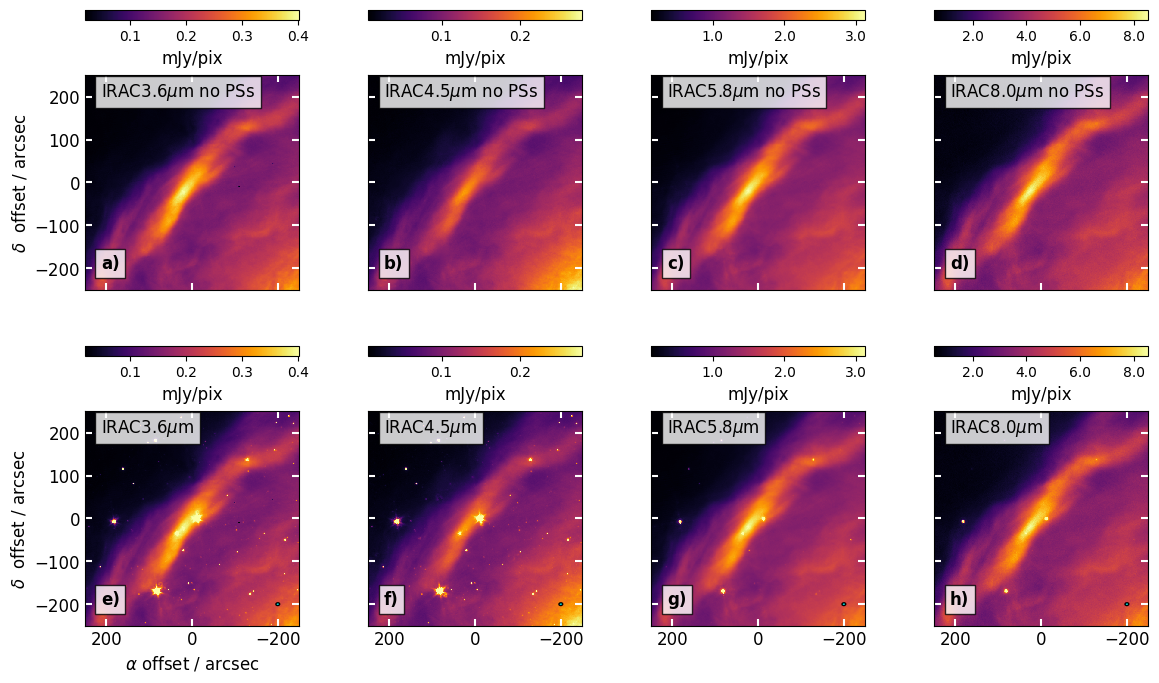}
   \caption{Comparison between the original IRAC images with the result of point-source subtraction.}
  \label{fig:IRstars}
\end{figure*}

Star subtraction was performed using {\tt
  StarXTerminator}\footnote{\url{https://www.rc-astro.com/software/sxt/}},
an AI-based tool that employs neural networks to remove stellar
sources from astronomical images. This tool was used as a plug-in
within the {\tt PixInsight}\footnote{\url{https://pixinsight.com/}}
image-processing platform. To assess whether the subtraction
introduced any bias, we masked the locations of the stars and
correlated the star-subtracted template with the original image,
finding a correlation slope of unity, which indicates that the
subtraction is robust. An example comparison of the result of
point-source subtraction is shown in Fig.\,\ref{fig:IRstars}.

\end{appendix}

\end{document}